# Joint Access-Backhaul Perspective on Mobility Management in 5G Networks

Rakibul Islam Rony, Akshay Jain, Elena Lopez-Aguilera, Eduard Garcia-Villegas and Ilker Demirkol
Dept. of Network Engineering
Universitat Politècnica de Catalunya
Barcelona, ES 08034
Email: {rakibul.islam.rony, akshay.jain}@upc.edu, {elopez, eduardg, ilker.demirkol}@entel.upc.edu

*Abstract*—The ongoing efforts in the research development and standardization of 5G, by both industry and academia, have resulted in the identification of enablers (Software Defined Networks, Network Function Virtualization, Distributed Mobility Management, etc.) and critical areas (Mobility management, Interference management, Joint access-backhaul mechanisms, etc.) that will help achieve the 5G objectives. During these efforts, it has also been identified that the 5G networks due to their high degree of heterogeneity, high QoS demand and the inevitable density (both in terms of access points and users), will need to have efficient joint backhaul and access mechanisms as well as enhanced mobility management mechanisms in order to be effective, efficient and ubiquitous. Therefore, in this paper we first provide a discussion on the evolution of the backhaul scenario, and the necessity for joint access and backhaul optimization. Subsequently, and since mobility management mechanisms can entail the availability, reliability and heterogeneity of the future backhaul/fronthaul networks as parameters in determining the most optimal solution for a given context, a study with regards to the effect of future backhaul/fronthaul scenarios on the design and implementation of mobility management solutions in 5G networks has been performed.

## I. INTRODUCTION

An expected multi-fold growth in data traffic and number of users [1], coupled with near static revenues and prohibitively high Capital Expenditures (CAPEX) and Operating Expenditures (OPEX) have prompted the wireless communities, both academic and industrial, to work towards a new generation of wireless technology, i.e., 5G. Given the exponential growth in traffic and users, future network scenarios are envisioned to be highly heterogeneous and dense. Therefore, with a vision to have a standard that caters to the aforementioned scenarios as well as to streamline the design, development and standardization efforts, 3GPP and ITU have listed out certain expectations from the 5G networks [2], [3]. A summary of some of these expectations, listed as challenges in [4], are provided in Table I.

From Table I it can be inferred that, in order to fulfill the expectations, new and innovative network architecture and resource management mechanisms are required. In addition to the innovative network architecture, current research efforts [5]–[8] have led to the identification of techniques such as Software Defined Networks (SDN), Network Function Virtualization (NFV), Distributed Mobility Management (DMM), Device-to-device (D2D) communications, etc., as being the pillars of 5G wireless networks. However, it is widely considered that, apart from the aforementioned enablers, mechanisms such as Mobility Management (MM), joint access-backhaul resource management, etc., will also play a significant role in realizing the 5G network objectives.

TABLE I
EXPECTATIONS FROM THE 5G FRAMEWORK

| Parameters | Support |
|---|---|
| Data Rates | 10-100x more than LTE data rates |
| Mobility | Support for high speed users (~500km/h) |
| Heterogeneous Networks | Mobility support in heterogeneous Radio Access Technologies along with multi-connectivity capabilities |
| CAPEX/OPEX | Sustainable |
| New deployment capabilities | Easy |
| Wireless device density | Support for 10-100x more devices |
| End-to-End latency | <1 ms |
| Quality of Experience (QoE) | Context based (flow, mobility profile, etc.) |
| Energy efficiency | High |

Network is transitioning towards a denser configuration (Fig. 1), leading to more challenging interference management, network design and efficient resource utilization problems. However, joint operation provides a promising solution, making the network more flexible, effective and resource efficient [9]. In this approach, access and backhaul networks can be integrated together, hence, allowing resource pooling, inter-dependency and efficient cooperation.

Next, mobility management, which is mostly agnostic to the backhaul network, enables the smooth handover of Mobile Node (MN) and its associated traffic in the event the MN switches its current point of attachment. However, for 5G networks, mobility management frameworks will need to cater to the highly dense and heterogeneous environments that will be prevalent. Consequently, the MM mechanisms in such a network will be susceptible to the backhaul network constraints, which will be unavoidable due to the joint design of access and backhaul.

Henceforth, in this paper a thorough study into 5G backhaul networks and its interconnect with mobility management is provided. To the best of our knowledge, such a discussion with regards to the effects of joint access-backhaul mechanisms on mobility management is unique. With this background, the rest of the paper is structured as follows: Section II provides a detailed study into 5G backhaul scenarios, requirements and potential backhaul solutions. Section III then briefly discusses the joint design of access and backhaul networks with different approaches. Following this discussion, Section IV presents the dependency and joint operation of MM with the backhaul network, which is often not considered. Finally, this paper is concluded with future research directions and corresponding challenges.

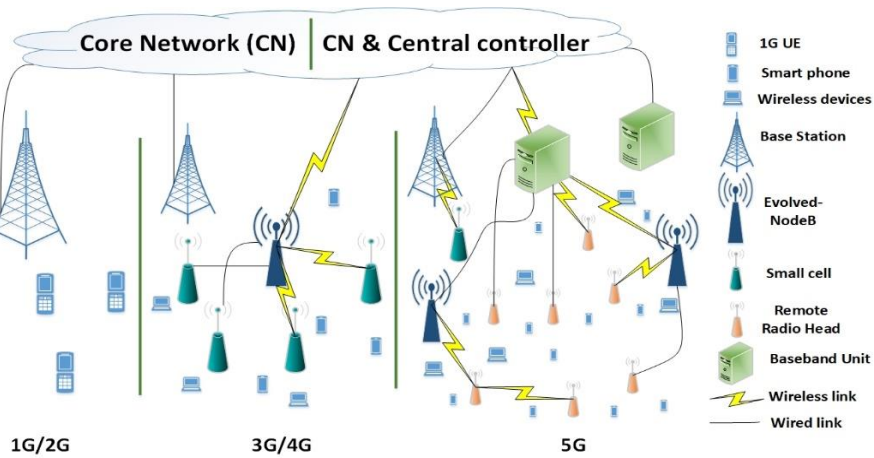

Fig. 1. Radio Access Network evolution

## II. 5G Backhaul Scenarios

Older generations, such as 1G and 2G were deployed using leased line, copper or fibre line as backhaul. Public Switched Telephone Network (PSTN) lines have also been considered as an option in few cases. Voice traffic in 1G and 2G was simply supported by backhaul links, which evolved from a collection of Time Division Multiplexed (TDM) links. Later, in 2G and 3G, microwave wireless links have also worked as backhaul links while backbone of the network was still wireline based [10]. However, due to multiple different use cases and deployment scenarios in future networks, solo wire-line based backhaul network is not a cost efficient option for the operators anymore. For cost efficient and fast deployment, wireless backhaul options are very attractive. Contradictory to their advantages, wireless backhaul links add interference in the network and have capacity and distance limitations. To take the advantages of both the aforementioned solutions, i.e., wired and wireless, 5G networks are anticipated to be deployed with heterogeneous backhaul networks. From the architectural point of view, 5G transport network is expected to be very complex and composed with backhaul, midhaul and fronthaul.

### Backhaul

Traditionally, the links connecting Base Station (BSs)/Evolved-NodeB (eNBs) (performing RAN processing) to the core network are called backhaul links (BH), which consist already in a popular term. In this scenario, links connecting one BS/eNB to another BS/eNB are also considered as BH. On the other hand, in the centralized approach, i.e. Centralized RAN (CRAN), the link connecting Baseband Unit (BBU) and the core network is also called BH. Moreover, in 5G networks, both Distributed (RAN processing is distributed to BSs) RAN (DRAN), and CRAN will co-exist and, in both cases, BH is carrying large amount of traffic to/from the core network. For a cost effective deployment, all these BSs can be connected to the core network and, thus, those BSs are linked to each other via the core network, although adding latency. According to [11], copper wire and wireless links can be used as BH links where optical fibre has not been already deployed. Different approaches such as Resilient Packet Ring (RPR), Optical Add Drop Multiplexing (OADM) ring technology and Wavelength Division Multiplexing (WDM) can be used for better performing BH with lower latency.

### Fronthaul

CRAN approach centralizes most of the RAN functionalities in BBU and the connecting links between BBU and Remote Radio Heads (RRH) are known as fronthaul (FH). The links connecting the RRHs to each other are also considered as FH link. Common Public Radio Interface (CPRI), Open Base Station Architecture (OBSAI) and Open Radio equipment Interface (ORI) are popular options for FH, although FH might have both wired and wireless links deployed. FH has already been justified as a key element of future networks having stringent requirements. Few novel interfaces for FH are also being explored such as, fronthaul-lite, Next Generation Fronthaul Interface (NGFI) and xHaul.

### Midhaul

In 3GPP terminology, the X2 based inter eNB interface is called the midhaul. However, with regards to future 5G

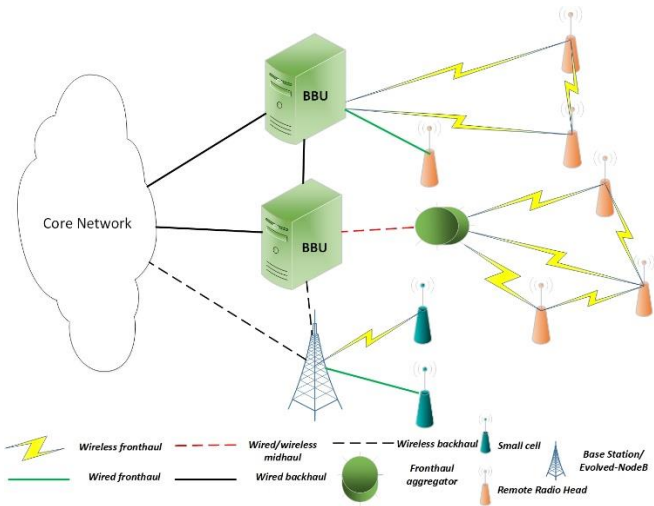

Fig. 2. Heterogeneous backhaul in 5G networks

based mobile networks, in [12] this term is used differently. According to [12], midhaul is the connecting link between aggregated fronthaul point and BBU. The idea is to benefit from multiplexing gains. Additionally, some data compression techniques can also be adopted in the aggregator to relax the requirements for the subsequent transport network. The midhaul links can be wired or wireless links according to network requirements and availability.

Fig. 2 helps to understand the separation between fronthaul, midhaul and backhaul. In subsequent sections, the term backhaul is used to refer to the entire transport network, including both fronthaul and midhaul. In few cases they might be mentioned separately for better understanding.

The complex heterogeneous transport network depicted in Fig. 2 will be a dominant element in 5G networks, which also needs to assure very high Quality of Service (QoS). With the evolution of RAN technologies, the expectations and popularity of mobile broadband access are growing by multiple fold. According to [2], International Mobile Telecommunications for 2020 (IMT-2020) is expected to support a connection density up to $10^6/\text{km}^2$. Moreover, the evolving (e.g. Carrier Aggregation (CA), HetNet) and disruptive features (e.g. Multi Radio Access Technology (RAT), multi-tenancy, enhanced mobility, etc.) of 5G require an exclusive transport network which is flexible and scalable. Moreover, to support the anticipated traffic, 5G networks are expected to employ the idea of frequency reuse as a promising solution. As a consequence, future mobile networks will utilize the concept of multiple sizes of small cells (e.g. atto-cells, femto-cells, pico-cells) in the network connecting very large amount of devices expecting high data rates. This deployment scenario with huge number of Access Points (APs) and users in the network introduces us to the concept of Ultra Dense Network (UDN). To support the UDN in the access network, a high capacity, low latency backhaul network is necessary to ensure that backhaul is not acting as bottleneck. Moreover, CRAN introduces more stringent requirements in the backhaul network, due to the fact that proper communication between RRH and BBU and between BBU and core network are needed, whereas a large amount of data needs to be transported in a very small amount of time. Nonetheless, apart from capacity and latency, enhanced synchronization is also a key requirement for 5G backhaul network.

To meet these aforementioned requirements, there are some already popular wired and wireless technologies being considered as backhaul solutions for 5G networks. All of them have their own advantages and shortcomings. Optical fibre as wireline backhaul is by far the best option in terms of capacity, latency and QoS, though it has shortcomings as less scalability and high deployment cost. Passive Optical Network (PON) technology for fibre has been evolving throughout the years, improving the performance of fibre based solutions. Besides, copper base wired backhaul with Asymmetric Digital Subscriber Line (ADSL), Very high speed Digital Subscriber Line (VDSL) and G.Fast technologies also provide promising wireline backhaul solutions with high link capacity, yet again not suitable for many use cases of access points. On the other hand, higher frequency wireless options provide larger link capacity, but they are very vulnerable to environmental effects. For instance, mmWave operating in three different bands, 60 GHz (V-band), 70/80 GHz and 90 GHz (E-band > 60 GHz) has recently begun to appear as attractive option for future wireless backhaul and access network technology, as it offers very large capacity (up to 10Gbps) compared to other wireless options. Moreover, advanced technologies, such as spatial multiplexing and beamforming can improve the overall performance of mmWave. Besides mmWave, already popular wireless options, such as Free Space Optical (FSO), Sub-6GHz, traditional microwave (Point-to-Point (PtP), Pointto-multi-Point(PtmP)), can also be considered as wireless backhaul options for 5G according to their availability and particular requirements. However, most of them require Line of Sight (LoS) propagation for reaching the expected performance. Fig. 3 illustrates different backhaul solutions.

### III. JOINT DESIGN OF ACCESS AND BACKHAUL IN 5G

In 5G, the transport network composed by BH and FH is expected to be a costly component, because of its heterogeneity, complexity and stringent requirements. Previously, including 3GPP architecture, radio access designs considered backhaul network to be sufficient [13], which is certainly not the scenario in upcoming 5G networks. In this situation, resource sharing and joint design of access and backhaul network can minimize the network CAPEX and OPEX significantly. Moreover, in dense networks APs have to serve one User Equipment (UE) cooperatively as several APs will be available for one UE. This cooperation needs to take into account the backhaul condition for the APs, best path (link quality, number of hops, etc.) for the UE and the access network conditions all together. To perform this cooperative operation in a cost efficient way, joint operation between access and backhaul, hence blurring the separation line between access and backhaul networks becomes necessary [13].

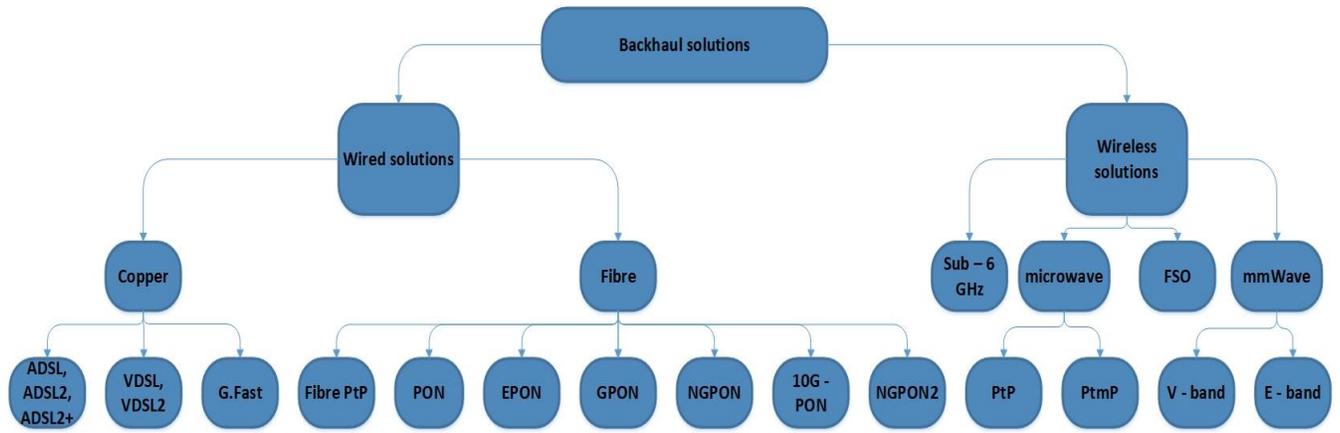

Fig. 3. Distinct Backhaul solutions

Additionally, as the wireless backhaul options are providing more attrac- tive solutions (i.e. cost efficiency, deployment feasibility, fast and easy deployment), technologies using the same resources (e.g. frequency channels) may be used by both access and backhaul networks. Therefore, in future networks the access and backhaul networks cannot be seen as separate entities, rather, integrated together to ensure the best use of resources [9], while solo optimization of access network is not enough anymore.

With this in mind, Fig. 4 depicts some different approaches to validate the Joint Design of Access and Backhaul (JDAB), where access and backhaul networks take into consideration each other requirements and availability. **Flexible RAN** allows the RAN functionalities to transition between CRAN and DRAN architectures on demand. Flexible-RAN is an idea to find out the trade-off between CRAN and DRAN according to the backhaul link quality and availability allowing the benefits of both approaches. **Functional split** in different layers (e.g. PHY layer, MAC layer) allows splitting of functions within a layer to achieve the centralization gain and to relax the BH requirements. For instance, if good capacity BH link is available, more functions can be centralized and vice-versa. **Access and BH awareness** is a context aware approach, where access and backhaul networks are aware to each other's requirements and limitations. Traditionally, the APs in the network are fed with equal amount of BH resources, which sometimes results in misuse of them. Hence, when access and BH are aware of each other, efficient resource distribution can be beneficial. Joint **interference management** of access and BH networks is very essential for in-band backhaul solutions, where access and backhaul networks use the same band for transmission. Under these situations, access and BH might act as interferer to each other and hence, joint management of interference is required. **Load balancing** is already a popular idea for balancing the load between different APs, however, traditional approaches do not consider the BH network load or congestion. Joint load balancing can take into account both the access and BH level load scenario and balance the load accordingly. **Resource allocation** schemes for the users in access network need to consider the BH resources and link quality along with access level resources. Similarly, resource allocation schemes for the BH networks should also consider the requirements in corresponding access networks. Joint **energy optimization** technique validates the idea of network wide energy optimization of both access and BH networks to increase the overall energy efficiency. For instance, to save energy, some access nodes might minimize the transmit power and few users will be handed over to a nearby AP without considering the BH situation of the new AP, which might create congestion. Whilst, network-wide energy optimization considers both access and BH performance degradation and tolerance level due to power minimization. Finally, similar to other mechanisms, traditional **MM** has also considered the backhaul network to be ideal or sufficient. In the subsequent section we mention the different parameters taken into account by different MM approaches and why BH network quality (i.e. link capacity and latency) must be considered.

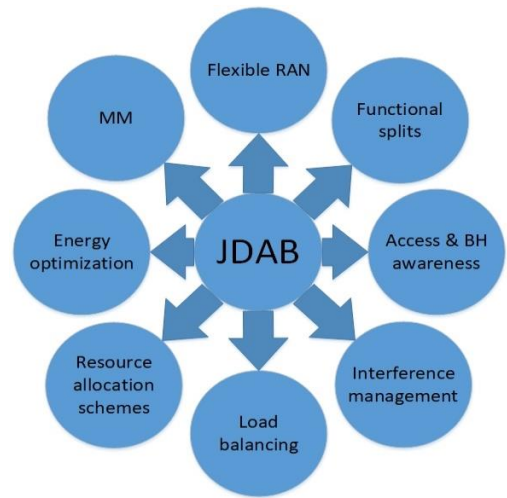

Fig. 4. Different approaches of joint Design of Access and backhaul

## IV. BACKHAUL RESOURCE CONSIDERATIONS IN MOBILITY MANAGEMENT

For any wireless technology to be ubiquitous and hence, successful, managing mobility of users is a critical aspect. By ensuring mobility in the network, the QoE for the users is enhanced. Consequently, standard bodies such as IETF, IEEE, 3GPP and ITU have over time proposed many standards and protocols that provide such mobility, whilst ensuring the requested QoS as well as maintaining fair utilization of network resources. Methods such as IEEE 802.21/802.21c [14], [15] provide mobility not just within a particular RAT but, they guarantee mobility to the user amongst various RATs (such as IEEE 802.11/16/15, etc.). Concurrently, protocols such as Mobile IPv6 (MIPv6), as proposed by IETF, help in ensuring continuity of service during network level mobility events. MiPv6 variants such as PMiPv6, which is a network centric approach (i.e., it does not need the user to participate in MM signaling), have found acceptance in 3GPP-LTE networks. Additionally, 3GPP, for LTE networks, has also defined the X2 and S1 handover procedures [16], depending on the type of interfaces present in a particular geographical area.

Whilst, the aforementioned strategies have sufficed the needs of the current day networks, 5G networks have been envisioned to be more dense, heterogeneous and dynamic. The extremely intricate and challenging nature of the 5G network, is illustrated through Fig. 1. It can be seen that the future scenarios will have higher density of users as well as of access points. Further, the various access points within the network, utilizing different RATs, will contribute to the heterogeneity of the network. In addition to the increased density and heterogeneity of the future network scenarios, as compared to the current scenario, the ability to deal with multiple mobility profiles, ranging from static sensors to fast moving users, will also be a critical component.

With this background, recent research efforts on designing mobility management solutions for 5G networks, such as [17]–[19], have proposed approaches which essentially equip the network/user with efficient RAT selection (handover management) methods or SDN based algorithms for fast path switch- ing and reduced latency during network level mobility events. These two broad classes of mobility management approaches focus on parameters such as network load, Received Signal Strength Indicator (RSSI), signaling cost, packet forwarding cost, user/network policies, etc. It is important to note here that the design and development of such mobility management techniques assume that the BH resources are uniform and unconstrained. However, the BH resources in 5G networks will be heterogeneous and non-ideal in nature (Section II). Consequently, the mobility management strategies for the future networks, in addition to the standard MM parameters, also need to take into consideration the uniqueness in the BH scenarios that will be prevalent in the 5G networks.

BH networks in the future 5G will be composed of both the wired and wireless media (Section II). Given the ultra- dense nature of the future networks with regard to the number of users accessing the network, the amount of available BH capacity becomes a critical factor whenever a handover decision is made. The critical nature of this factor can be further understood from the fact that, if multiple users are assigned the same access point and, thus, the same BH resources (through handovers of initial attach procedures), then the probability that a particular BH link is congested will be high.

It is important to state here that, in addition to the user requested handovers, traffic transfer decisions in order to perform load balancing tasks also implicitly invoke mobility management protocols. In such scenarios, the critical nature of BH resources should also be taken into consideration.

Not only will BH networks be heterogeneous, but they will also serve APs which are either connected directly to the BBU pool or can be reached via multiple-hops (Fig. 1). Further, it has been discussed that the available link capacity will be an important factor for consideration whilst generating a mobility management decision. However, there might be scenarios where the users are assigned to APs which are connected to the core network via multiple hops. In such a scenario, if the users are accessing delay sensitive services, then such an AP assignment will most certainly lead to a degradation in the QoS as the BH induced network latency will increase.

And hence, in addition to the available BH link capacity, the number of hops that the user will have to traverse to reach the core network, after the handover process, will also be critical in devising mobility management strategies for 5G networks. Additional to the aforesaid factors, the availability (there might be scenarios where a particular link is congested or non-functional) and reliability (wired links are always more reliable than wireless links; amongst wireless links, Signal-to-noise Ratio (SNR) of each individual link will be a crucial determining factor towards their relative reliability) of the BH resources also merit consideration by the mobility management strategies employed in the 5G networks. Therefore, in order to generate an optimal mobility management solution that also takes into account the unique scenario that the future BH networks will present, the MMaaS paradigm [20], which aims to provide flexibility to MM mechanisms through its provision of granularity in service, will need to incorporate all the BH network related factors discussed so far. Consequently, the flexibility and granularity offered by the MMaaS paradigm will be potentially enhanced further.

To illustrate, from Fig. 5, it can be seen that the mobility management services are employed as an application on top of the SDN-controller (SDN-C). These services, which maybe present on a cloud, have the complete network view. Subsequently, parameters from both the user as well as the network can be extracted by the aforementioned MM application. The extracted parameter values consist of detected access point SINR/RSSI/SNR value, flow types, mobility profile, user/network policies, BH network scenario information, etc. After these parameters are analyzed, a mobility management solution based on the current user and network context is generated, which is then executed by the SDN controller. It is important to note that, with the BH network scenario informa-tion, the network can provide improved mobility management solutions. Thus, the already available granularity perspectives in the MMaaS paradigm, i.e., flow, mobility profile, network load, user/network policies [20], can now utilize this extra access and BH network information to customize the offered MM solutions even further.

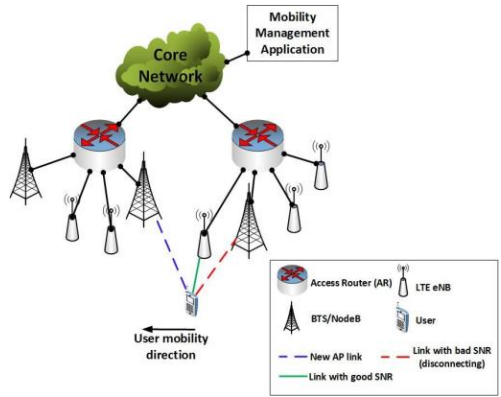

Fig. 5. Mobility Management scenario in future networks.

For example, if a user has two flows, i.e., a delay-sensitive and a delay tolerant flow, then whilst handing over, and assuming there are multiple APs in the vicinity and the user has multi-connectivity capabilities, the delay-sensitive flow can be associated with an AP that is connected to the BH with a link of sufficient capacity, as well as is employing less number of hops, so as to maintain the QoS. But, the delay-tolerant flow, for which the QoS requirements are not very high, can be assigned an AP which needs more hops as well as lesser amount of available capacity in the backhaul.

Whilst, the aforementioned mechanism provides a simple solution to the challenge of utilizing BH information for MM there can be scenarios where the shorter route (i.e., with less number of hops) is congested. Consequently, in those scenarios, a longer route (more hops) can meet the specified latency requirements, and hence the requested QoS, as compared to the former. Further, heterogeneity as has been stressed before, needs to be taken into consideration given that propagation and transmission delays can vary depending on the technology being used. From the above discussion, it is clear that the BH scenario information can vastly enhance the capabilities of the MMaaS paradigm, and consequently, the mobility management services that will be offered in the 5G networks.

## V. Conclusion

Joint design of access and backhaul networks creates new opportunities in the design of 5G networks. This paradigm can ensure the best usage of precious resources, which makes 5G networks more cost and resource efficient. Moreover, this joint design opens the opportunity for MM to be more backhaul dependent, realistic and effective. In this paper, we put together the different approaches where access and backhaul can be interdependent, and try to justify MM as an essential part of this approach. Hence, intelligent algorithm is required for MM, where backhaul dependency and access network quality both are taken into account. Finally, some of the mentioned backhaul options are not fully developed, yet suffer from high cost and shortage of bandwidth, and so, 5G backahul options require more in-depth studies.

## Acknowledgment

This work has been supported in part by the EU Hori- zon 2020 research and innovation programme under grant agreement No. 675806 (5GAuRA), and by the ERDF and the Spanish Government through project TEC2016-79988-P, AEI/FEDER, UE.